\documentclass[%
 reprint,
 superscriptaddress,
 amsmath,amssymb,
 aps,
 pra,
 floatfix,
]{revtex4-1}

\usepackage{graphicx}
\usepackage{dcolumn}
\usepackage{bm}
\usepackage[sort&compress]{natbib}
\usepackage{xcolor}
\usepackage{cleveref}
\usepackage{mathtools}  
\usepackage{soul}

\usepackage{braket}

\Crefname{equation}{Eq.}{Eqs.}
\crefname{equation}{Eq.}{Eqs.}
\Crefname{figure}{Fig.}{Figs.}
\crefname{figure}{Fig.}{Figs.}
\Crefname{appendix}{Appendix}{Appendices}
\crefname{appendix}{Appendix}{Appendices}
\Crefname{section}{Sec.}{Secs.}
\crefname{section}{Sec.}{Secs.}

\newcommand{\id}{\mathbb{I}}
\newcommand{\pr}{\mathbb{P}}
\newcommand{\argmin}{\mathrm{argmin}}
\newcommand{\argmax}{\mathrm{argmax}}
\newcommand{\qsbo}{QSBO}

\begin{document}

\title{Quantum-Enhanced Simulation-Based Optimization}

\author{Julien Gacon}
\affiliation{IBM Research -- Zurich}
\affiliation{ETH Zurich}

\author{Christa Zoufal}
\affiliation{IBM Research -- Zurich}
\affiliation{ETH Zurich}

\author{Stefan Woerner}
\email{wor@zurich.ibm.com}
\affiliation{IBM Research -- Zurich}

\date{\today}

\begin{abstract} 
In this paper, we introduce a quantum-enhanced algorithm for simulation-based optimization. Simulation-based optimization seeks to optimize an objective function that is computationally expensive to evaluate exactly, and thus, is approximated via simulation. Quantum Amplitude Estimation (QAE) can achieve a quadratic speed-up over classical Monte Carlo simulation.
Hence, in many cases, it can achieve a speed-up for simulation-based optimization as well. Combining QAE with ideas from quantum optimization, we show how this can be used not only for continuous but also for discrete optimization problems. Furthermore, the algorithm is demonstrated on illustrative problems such as portfolio optimization with a Value at Risk constraint and inventory management.
\end{abstract}

\maketitle

\section{\label{sec:introduction}Introduction}

In mathematical optimization, the goal is to find values for decision variables such that a given objective function is optimized.
If the evaluation of the objective function involves simulation, the problem is referred to as simulation-based optimization (SBO) \cite{Gosavi1997}. 
Simulation is used to evaluate a function that is computationally too expensive to be evaluated analytically, e.g., because a system of interest is too complex or because it involves uncertainty.
Especially in the latter case, Monte Carlo simulation is popular since the estimation error scales as $\mathcal{O}(M^{-1/2})$, where $M$ denotes the number of function evaluations, and is independent of the dimension of the system \cite{Morton2001}.

Quantum Amplitude Estimation (QAE) \cite{Brassard2000} is a quantum algorithm that provides a quadratic speedup over classical Monte Carlo simulation, i.e., its estimation error scales as $\mathcal{O}(M^{-1})$. 
Like Monte Carlo simulation, QAE can be applied to a large variety of problems. Recent research has already investigated the applicability of this algorithm to several tasks, for instance, risk analysis \cite{Woerner2018, Egger2019} and option pricing 
\cite{Rebentrost2018, Stamatopoulos2019, Vazquez2020}.

In the context of optimization, other quantum algorithms have already been investigated, such as the Variational Quantum Eigensolver (VQE) or Quantum Approximate Optimization Algorithm (QAOA) \cite{Peruzzo2014, Farhi2014, Moll2017, Barkoutsos2019}.
These methods are applicable to Quadratic Unconstrained Binary Optimization (QUBO), where the problem can be mapped to an Ising Hamiltonian.
However, unlike the algorithm presented here, 
they cannot be applied to other classes of optimization problems.

Within this paper, we show how QAE can be used to accelerate SBO over both continuous and discrete variables and introduce a quantum-enhanced SBO algorithm, to which we shall refer to as \qsbo.
\qsbo{} is particularly suited for objective functions that are defined as expectation value, variance, cumulative distribution function (CDF) or (conditional) value at risk ((C)VaR) of a random variable and functions thereof.
Notably, all these measures can be interpreted as expectation value, c.f.~\cref{app:cdf,app:var}.

The algorithm is applied to small instances of practically relevant problems from inventory management and finance, more precisely, the newsvendor 
problem and portfolio optimization with a VaR-based objective function. 
The implementations are based on Qiskit \cite{Qiskit}.

The remainder of the paper is structured as follows. 
In \cref{sec:preliminaries}, we discuss the QAE algorithm and how it can accelerate the estimation of expected values.
\Cref{sec:method} describes in detail our \qsbo{} algorithm. 
Applications of the proposed algorithm are presented in \cref{sec:applications}. Finally, \cref{sec:discussion} concludes this work and discusses possible extensions.

\section{\label{sec:preliminaries} Quantum Amplitude Estimation}

Given a unitary operator $\mathcal A$ on $n$ qubits, such that
\begin{equation*}
    \mathcal{A} \ket{0} = \sqrt{1 - a} \ket{\Psi_0} + \sqrt{a} \ket{\Psi_1},
\end{equation*}
where $a \in [0, 1]$ and $\ket{\Psi_0}$ and $\ket{\Psi_1}$ are two normalized and orthogonal states, then QAE computes an estimate $\tilde a$ of $a$. 

\begin{figure}
    \centering
    \includegraphics[width=\linewidth]{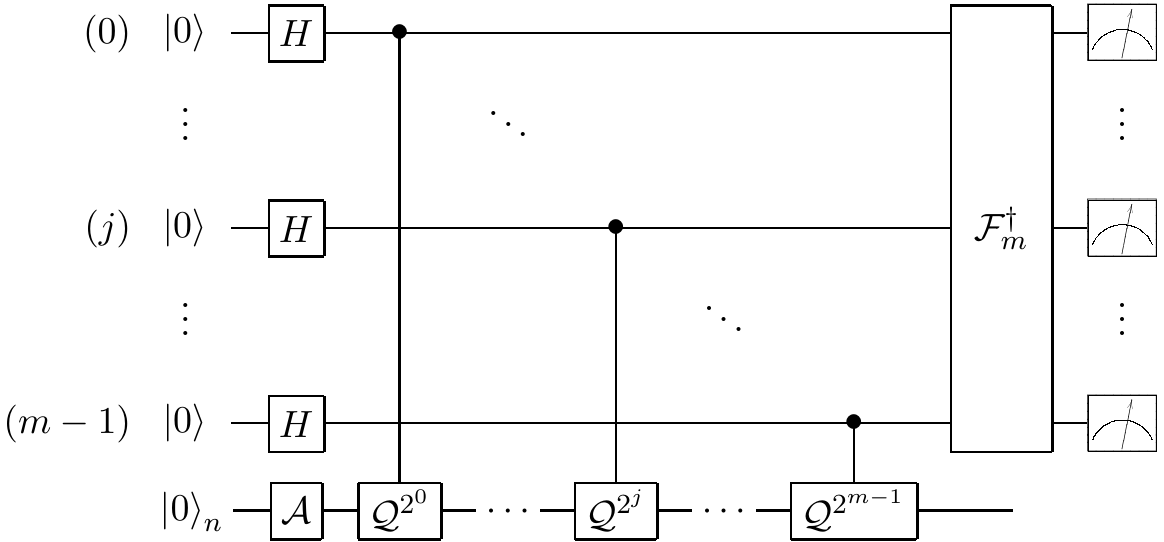}
    \caption{The circuit for the canonical QAE algorithm. The top $m$ qubits, on which the Hadamard
             gates $H$ are applied, are the evaluation qubits.
             The bottom $n$ qubits represent the states $\ket{\Psi_0}$ and $\ket{\Psi_1}$. 
             The operation $\mathcal{F}^\dagger_m$ is the inverse QFT. The evaluation qubits are measured in the computational basis
             and are mapped onto an estimate $\tilde{a}$ of $a$.}
    \label{fig:qae_circuit}
\end{figure}

Canonical QAE \cite{Brassard2000} is based on Quantum Phase Estimation (QPE) \cite{Kitaev1995}. 
First, the $n$ state qubits are initialized using $\mathcal{A}$. 
Then, $m$ evaluation qubits are added and initialized in the equal superposition state.
For the next step, an operator 
\begin{equation*}
\mathcal{Q} = \mathcal{A} \mathcal{S}_0 \mathcal{A}^{\dagger} \mathcal{S}_{\Psi_0}
\end{equation*}
is constructed, with the reflections
\begin{equation*}
\mathcal{S}_0 = \mathbb{I} - 2 \ket{0}\bra{0}  \text{ and }
\mathcal{S}_{\Psi_0} = \mathbb{I} - 2 \ket{\Psi_0}\bra{\Psi_0}, 
\end{equation*}
and $\mathbb{I}$ denoting the identity. The operator  $\mathcal{Q}$ is applied in total $M  - 1$ times, $M = 2^m$, controlled by the evaluation qubits. After an inverse quantum Fourier transform (QFT) on the $m$ evaluation qubits, these qubits are measured and the resulting integer is denoted by $y \in \{0, \ldots, M-1\}$.
Last, we set 
\begin{equation*}
\tilde{a} = \sin^2\left(\frac{y\pi}{M}\right),
\end{equation*}
where $|a - \tilde{a}| \leq \pi/M$ with probability at least $8/\pi^2$.
Due to the symmetry of sine, the algorithm output $\tilde{a}$ lies on a grid of $M/2+1$ possible values in $[0, 1]$.
The respective circuit for canonical QAE is illustrated in Fig.~\ref{fig:qae_circuit}.

The accuracy of the estimate $\tilde{a}$ is determined by the number of evaluation qubits $m$, and scales as $\mathcal{O}(M^{-1})$, while the circuit depth scales as $\mathcal{O}(M)$, i.e., the applications of $\mathcal{Q}$.
Compared to this, the standard deviation of the Monte Carlo estimator 
scales as $\mathcal O(M^{-1/2})$ for $M$ function calls, and thus, QAE achieves a quadratic speedup.
Since QAE is probabilistic, it must be repeated to obtain a good estimator with high confidence.
By executing the algorithm multiple times and taking the median of all estimates as the final result, the success probability quickly approaches one \cite{Woerner2018}.

Canonical QAE results in very complex circuits and provides only a discrete resolution for the estimate $\tilde{a}$ based on the $m$ evaluation qubits.
Thus, different variants of QAE were recently proposed that improve QAE in terms of accuracy and complexity \cite{Wie2019,Suzuki2019,Aaronson2019,Grinko2019,Nakaji2020}.
These algorithms provide a continuous range of 
estimates and can significantly reduce circuit width and 
depth by removing the ancilla qubits as well as the QFT.
Especially due to the continuous range for estimated values, these QAE variants are better suited for SBO as illustrated in \cref{sec:method}.

The remainder of this section shows how to use QAE to estimate an expected value $\mathbb{E}[f(X)]$, for a given random variable $X$ and function $f: \mathbb{R} \rightarrow [0, 1]$. Expected values in this form commonly appear as objective function in SBO.

Suppose $f: \mathbb{R} \mapsto [0, 1]$ and a corresponding unitary
operator $F$ given by
\begin{equation}
    F\ket{x} \ket{0} = \sqrt{1 - f(x)} \ket{x} \ket{0} + \sqrt{f(x)} \ket{x} \ket{1}.
    \label{eq:f_on_amplitudes}
\end{equation}
Such an operator $F$ can be constructed, for instance, using quantum arithmetic and related techniques \cite{Haener2018, Vazquez2020}.
In this work, we resort to the approach presented in \cite{Woerner2018},
which approximates $f$ using a Taylor expansion and uses 
controlled Pauli-rotations to rotate the function values onto the qubit amplitudes.
This mapping allows to trade off precision against circuit complexity by choosing the number of Taylor terms in the approximation of $f$ and circumvents the need for quantum arithmetic.

Next, suppose a discrete random variable $X$ taking values in $\Omega_X = \{x_i\}_{i=0}^{N-1}$, $N = 2^n$, for a given $n$, with corresponding probabilities $p_{x_i} = \pr[X = x_i]$.
Then, the expectation value can be written as
\begin{equation}
    \mathbb E\left[f(X)\right] = \sum_{i=0}^{N - 1} p_{x_i} f(x_i)
       = \sum_{\hat{x}=0}^{N - 1} p_{\phi(\hat{x})} f(\phi(\hat{x})),
    \label{eq:monte-carlo-estimate}
\end{equation}
where $\phi: \{0, \ldots, N - 1\} \rightarrow \Omega_X$ 
represents an affine transformation from $\hat{x} \in \{0, \ldots, N - 1\}$ to $x \in \Omega_X$.

Now, we can encode $\mathbb{E}[f(X)]$ in $\mathcal{A}$ using \cref{eq:f_on_amplitudes} and \cref{eq:monte-carlo-estimate}.
First, we load the discretized probabilities $p_x$ into the amplitudes of $n$ qubits by means 
of an operator $\mathcal P_X$ to construct the state
\begin{equation*}
    \mathcal{P}_X \ket{0}_n = \sum_{\hat{x}=0}^{N - 1} \sqrt{p_{\phi(\hat{x})}} \ket{\hat{x}}_n.
\end{equation*}
Preparing a qubit register to (approximately) represent a probability density function (PDF) can be done, e.g., by using quantum arithmetic if the function is efficiently integrable \cite{Grover2002}, with matrix products states for smooth and differentiable functions \cite{Holmes2020}, or by employing quantum generative adversarial networks to approximate generic functions \cite{Zoufal2019}.
Then, we add one more qubit and use $F$ to define $\mathcal{A} = F(\mathcal{P}_X \otimes \id)$.
The state after applying $\mathcal{A}$ is given by
\begin{equation}
\begin{aligned}
    \mathcal A \ket{0}_n\ket{0} &= \sum_{\hat{x}=0}^{N - 1} \sqrt{p_{\phi(\hat{x})}(1 - f(\phi(\hat{x})))} \ket{\hat{x}}_n \ket{0} \\
                                &+ \sum_{\hat{x}=0}^{N - 1} \sqrt{p_{\phi(\hat{x})} f(\phi(\hat{x}))} \ket{\hat{x}}_n \ket{1},
    \label{eq:expectation_value}
\end{aligned}
\end{equation}
where $\ket{\Psi_0}$, respectively $\ket{\Psi_1}$, correspond to the states where the last qubit is $\ket{0}$, respectively $\ket{1}$.
This implies that
\begin{equation*}
    a = \sum_{\hat{x}=0}^{N - 1} p_{\phi(\hat{x})} f(\phi(\hat{x}))
\end{equation*}
which is the desired expected value from \cref{eq:monte-carlo-estimate}.

QAE can not only be used to estimate expected values, but also the variance, cumulative distribution functions, and the (Conditional) Value at Risk \cite{Woerner2018}. This is discussed in more detail in \cref{app:cdf,app:var}.

\section{\label{sec:method}Quantum-enhanced simulation-based optimization}

Suppose a discrete random variable $X$ taking values in $\Omega_X$, a decision variable $y \in \mathbb{R}^d$, and a function $f: \Omega_X \times \mathbb{R}^d \rightarrow \mathbb{R}$.
Then, we are interested in finding
\begin{equation*}
    y^* \in \argmin_{y}~ \mathbb{E}[f(X, y)].
\end{equation*}
If we formulate the evaluation of this expectation value in terms of QAE, a classical optimization routine can be employed to find the optimal value $y^*$.
This decision variable $y$ may be continuous, e.g., the optimal price of a product
or the location of a sensor, or discrete, e.g., the optimal set of assets to include in
a portfolio or the optimal number of articles to have in stock.

\begin{figure}
    \centering
    \includegraphics[width=\linewidth]{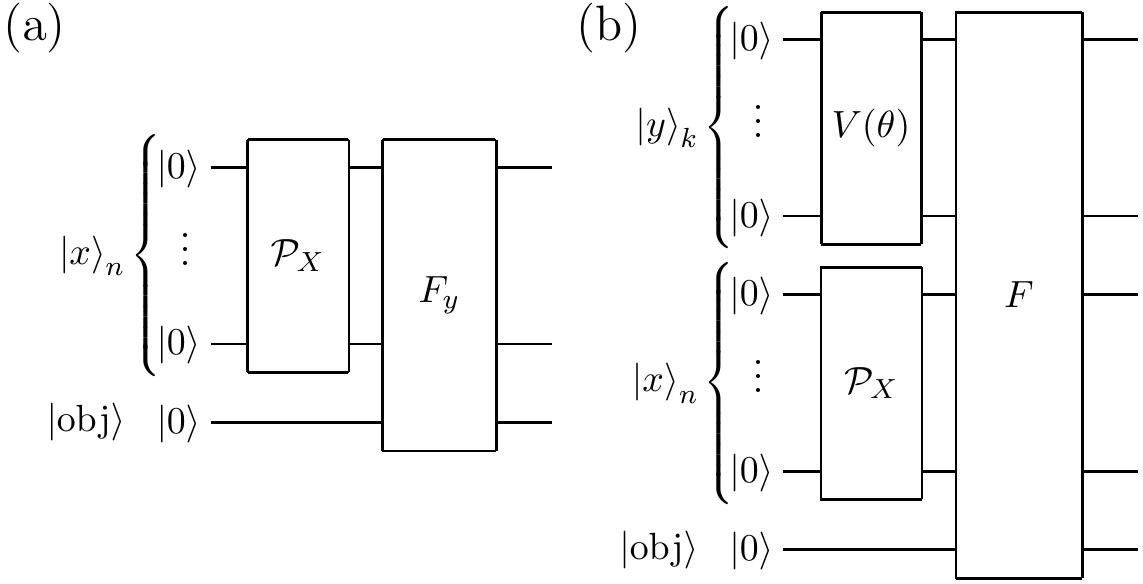}
    \caption{The circuit diagrams implementing the $\mathcal{A}$ operator. 
             If the decision variable $y$ is continuous it is a parameter of the function
             mapping $F_y$, see (a). In the discrete case, $y$ is represented using qubits and the state parameterized via the circuit $V(\theta)$, see (b).}
    \label{fig:a_operator}
\end{figure}

First, the continuous case is discussed.
If $y$ is continuous, then, the expectation value $\mathbb{E}[f(X, y)]$ can be evaluated by preparing 
the operator $\mathcal{A}$ as in \cref{eq:expectation_value}. 
The value of $y$ is handled as parameter of the function $f(X, y) = f_y(X)$ and, thus, 
of $F = F_y$, i.e.,
\begin{equation*}
    F_y\ket{x} \ket{0} = \sqrt{1 - f_y(x)} \ket{x} \ket{0} + \sqrt{f_y(x)} \ket{x} \ket{1}.
\end{equation*}
The action of $\mathcal{A}$, as shown in \cref{fig:a_operator}(a), can be formulated as \begin{equation*}
    \mathcal{A}_y \ket{0}_{n+1} = F_y(\mathcal{P}_X \otimes \id)\ket{0}_n \ket{0},
\end{equation*}
where the top $n$ qubits encode the random variable $X$ and 
the last qubit is used to apply $F_y$ and to mark the 
states $\ket{\Psi_0}$ and $\ket{\Psi_1}$.

A single evaluation of the objective consists of preparing the parameterized operator $\mathcal{A}_y$,
constructing $\mathcal{Q}_y$, and then, running 
QAE to obtain $\tilde{a}_y \approx \mathbb{E}[f(X, y)]$.
Using a numerical optimization routine with respect to the QAE output as objective function allows to update the parameter $y$ in order to find $y^*$.

The canonical formulation of QAE, as described in \cite{Brassard2000}, can be problematic for this optimization.
If $m$ evaluation qubits are used, the estimates of this QAE lie on a fixed grid of $M/2 + 1$ points, and thus, lead to a step function.
We illustrate this using an example in \cref{sec:toy}.
The resulting piece-wise constant approximation of the objective function can be challenging to optimize.
QAE-variants that provide continuous estimates \cite{Suzuki2019, Wie2019, Aaronson2019, Grinko2019, Nakaji2020} do not introduce such artificial discontinuities and are, therefore, advantageous in this context.

Suppose, now, that the decision variable $y$ is not continuous but discrete.
In this case, $y$ could still be treated as a parameter of $\mathcal{A}$, requiring a numerical optimizer that searches over discrete values. 
In this work, we focus on a different approach, motivated by VQE/QAOA, where $y$ is represented by qubit basis states. 

Given a discrete decision variable $y$ taking $2^k$ values, we represent it using $k$ qubits. For instance, if $y \in \{0, 1\}^k$, we map $y$ to $\ket{y_0, \ldots, y_{k-1}}_k$, although, other mappings are possible, e.g., to represent integer values.

The quantum state representing the decision variable $y$ is
parameterized, such that an optimization routine can tune the parameters, and, thereby, search a suitable solution $y^*$.
Inspired by VQE, this parameterization can be implemented with a parameterized circuit $V(\theta)$, $\theta \in \mathbb{R}^p$,
to construct a trial state $\ket{\psi(\theta)} = V(\theta)\ket{0}$.

The choice of $V$ determines the Hilbert space that can be scanned for the optimal
solution $\ket{y^*}_{k}$ and is, thus, crucial for implementing a successful scheme.
Ideally, $V$ can represent a large fraction of the respective Hilbert space while having low circuit depth, i.e.~a low number of non-parallel circuit operations.
In \cite{Sim2019}, parameterized circuits are compared with respect to their complexity and state representation capabilities.

While in VQE, this parameterized circuit is used as trial wave function for the ground state of a molecule, here, it is employed as trial state for the solution.
In other words, we define $\ket{y(\theta)}_k = V(\theta) \ket{0}_k$.
Then, the optimization problem can be formulated as
\begin{equation}
    \theta^* \in \argmin_{\theta}~ \mathbb{E}[f(X, y(\theta))],
\end{equation}
such that for a given $\theta$ the result of QAE is based on a superposition $\ket{y(\theta)}$ over possible solution candidates, i.e., $y(\theta)$ is a random variable as well, instead of a fixed value.
The discrete probability distribution of candidate values is given by the sampling probability, i.e.
\begin{equation*}
    \pr[y(\theta) = i] = \vert \braket{i|y(\theta)} \vert^2, \quad i \in \{0, \ldots, 2^k - 1\}.
\end{equation*}

Now, the $\mathcal{A}$ operator consists of $\mathcal{P}_X$ loading the probability distribution of the random variable $X$ into the qubit register $\ket{x}_n$, the trial state $V(\theta)\ket{0}_k$ representing the decision variable $y$ in register $\ket{y(\theta)}_k$ and a function mapping $F$ acting on $\ket{x}_n$ as well as $\ket{y(\theta)}_k$
\begin{equation*}
    F \ket{x}\ket{y}\ket{0} = \ket{x}\ket{y}
        \left(\sqrt{1 - f(x,y)}\ket{0} + \sqrt{f(x,y)} \ket{1}\right).
\end{equation*}
In total, this leads to
\begin{equation}
    \mathcal{A}_{\theta}\ket{0}_{k+n+1} = F (V(\theta) \otimes \mathcal{P}_X \otimes \id) \ket{0}_k \ket{0}_n \ket{0},
\end{equation}
see \cref{fig:a_operator}(b).

\section{\label{sec:applications}Applications}

This section presents numerical simulation results for various applications of the proposed \qsbo{} algorithm.
The numerical optimization routine employed in the classical optimization steps is the COBYLA optimizer \cite{Powell1994}.
Other choices are possible, but the comparison of different routines is beyond the scope of this work.

\begin{figure}
    \centering
    \includegraphics[width=0.8\linewidth]{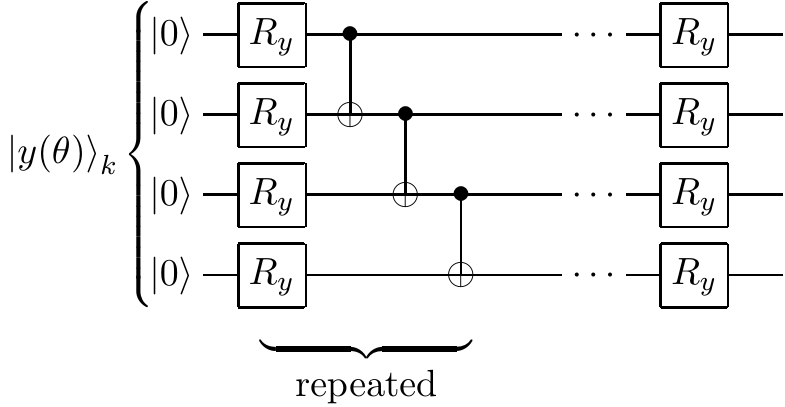}
    \caption{The $R_y$ trial state $V(\theta)\ket{0}_k$ is shown here on $k=4$ qubits. Every $R_y$ gate has one free parameter which can be optimized. The entanglement strategy in this circuit is referred to as linear.}
    \label{fig:ry}
\end{figure}

The trial state $V(\theta)\ket{0}$ for the discrete variables is given by a circuit 
with two alternating layers of parameterized $R_y$ rotations and linear CNOT entanglement, see \cref{fig:ry}.
Increasing the number of repetitions of the layers, leads to more parameters and, thus, to more degrees of freedom which typically gives access to a larger state space but is also more complex to optimize \cite{Sim2019}.

Furthermore, the probability density functions $\mathcal{P}_X$ of the random variables $X$ are loaded exactly into quantum states using the respective uncertainty models 
provided by Qiskit \cite{Qiskit}.

First, we consider an illustrative example with a quadratic objective function for both, continuous and discrete variables.
Then, we discuss the newsvendor problem as an exemplary inventory management problem. 
Finally, \qsbo{} is applied to portfolio optimization using an objective function based on expected return and VaR.

\subsection{\label{sec:toy}Quadratic objective function}

We consider the following continuous optimization problem 
\begin{equation}
    y^* = \argmin_{y}~ \mathbb E\left[(X - y)^2\right],
    \label{eq:toy_objective}
\end{equation}
where $X \sim \mathcal{N}(1, 1)$ truncated to $\Omega_X = [0, 2]$ and $y \in [0, 2]$.
We discretize $X$ using $n = 2$ qubits, i.e., $\mathcal{A}_y$ acts on three qubits in total.
The values of $X$ are discretized and represented in binary using the states of $n$ qubits $q_i$ 
to define $\hat{x} = \sum_{i=0}^{n - 1} 2^i q_i$ and $x = \phi(\hat{x})$.
The affine transformation $\phi: \{0, \ldots, 2^n - 1\} \rightarrow \Omega_X$, 
for $\Omega_X = [\ell, u]$, reads
\begin{equation*}
    \phi(\hat{x}) = \ell + (u - \ell) \frac{\hat{x}}{2^n - 1}.
\end{equation*}

\begin{figure}
    \centering
    \includegraphics[width=\linewidth]{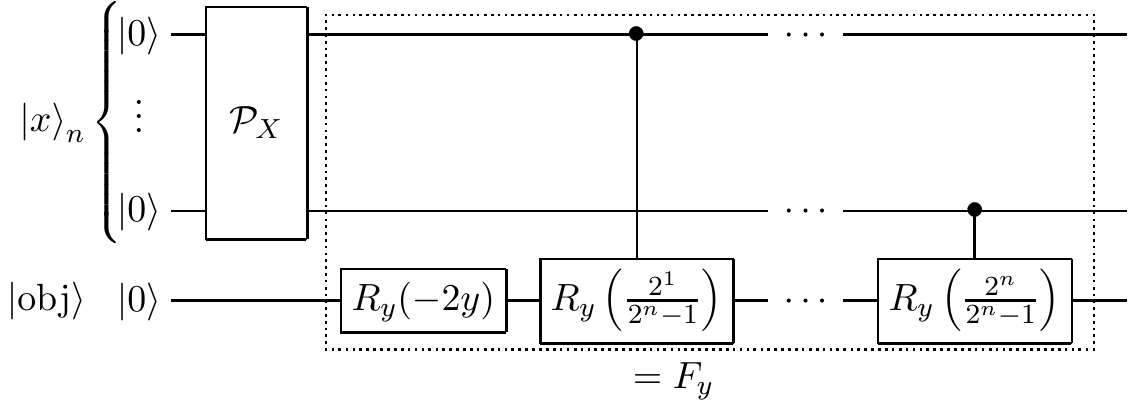}
    \caption{The circuit implementing $\mathcal{A}_y$ to compute $\mathbb{E}[(X - y)^2]$ for 
             a continuous parameter $y$.
             First, the probability distribution is loaded using $\mathcal{P}_X$.
             Then, the expression $\sin(c(x - y))$ is rotated onto the amplitude
             of state $\ket{1}$ of the bottom qubit using linearly controlled Pauli-$Y$ rotations.
             This is discussed in detail in \cite{Woerner2018}.
             }
    \label{fig:continuous_toy_example}
\end{figure}

Following the approximation introduced in \cite{Woerner2018}, this quadratic function can be evaluated as
\begin{equation*}
   (x - y)^2 = \frac{\sin^2(c(x - y))}{c^2} + \mathcal{O}(c^2),
\end{equation*}
where the factor $c > 0$ controls the accuracy of the approximation.
Accepting the introduced error $\mathcal{O}(c^2)$ and dropping $1/c^2$, this can easily be prepared using (controlled) $R_y$ gates.
The operator $\mathcal{A}_y$, then, acts as 
\begin{equation}
    \begin{aligned}
  \mathcal{A}_y \ket{0}_{n+1} &= \sum_{\hat{x}=0}^{2^n - 1} 
    \sqrt{p_{\phi(\hat{x})}} \sin\left(c\left(\phi(\hat{x})  - y \right)\right) \ket{\hat{x}}_n \ket 1 \\
                 &+ \sum_{\hat{x}=0}^{2^n - 1} \sqrt{p_{\phi(\hat{x})}} \cos\left(c\left(\phi(\hat{x})  - y \right)\right) \ket{\hat{x}}_n \ket 0,
    \end{aligned}
  \label{eq:continuous_toy_model_a_operator}
\end{equation}
and we can use it within QAE.
\Cref{fig:continuous_toy_example} presents the corresponding circuit.
The output of QAE is transformed to an estimate of the
expectation value by reverting the applied scaling,
\begin{align*}
  \mathbb E\left[(X - y)^2\right] \approx \frac{\tilde{a}}{c^2}.
\end{align*}

\begin{figure}
    \centering
    \includegraphics[width=0.9\linewidth]{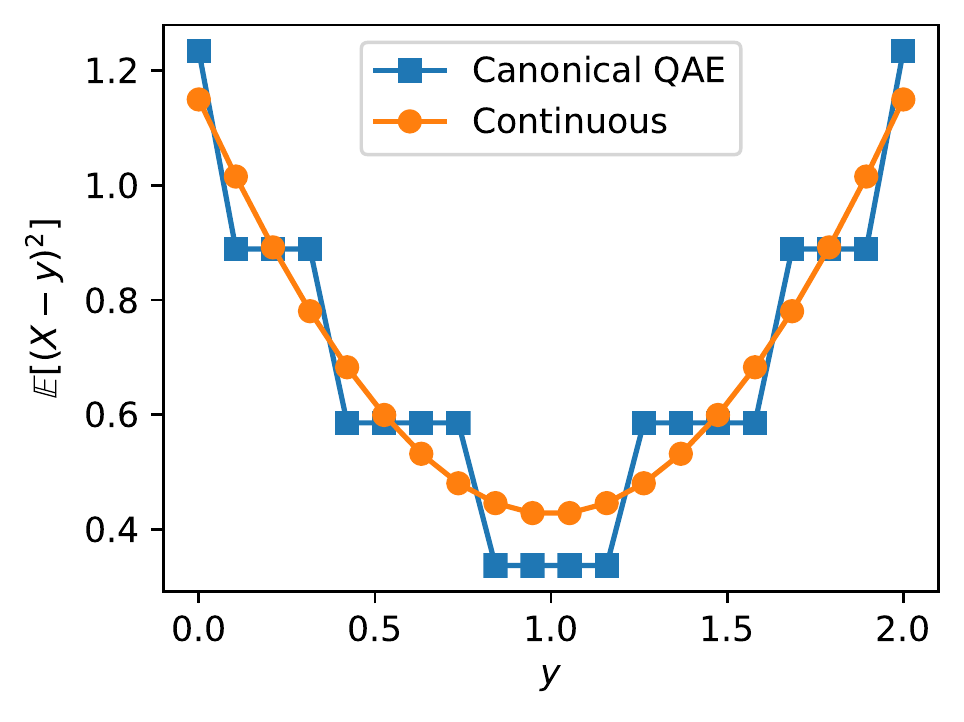}
    \caption{The evaluated objective function $\mathbb{E}[(X - y)^2]$
    for $X = \mathcal{N}(1, 1)$, truncated to $[0, 2]$ and different values of $y$. Using the canonical QAE algorithm (here, with $m=5$ evaluation qubits) for the evaluation leads to a step
    function (blue squares) which can be challenging to optimize in more difficult examples.
    The QAE variant using maximum-likelihood post-processing has access to
    continuous estimates and mirrors the smoothness of the objective function.
    }
    \label{fig:discrete_vs_continuous}
\end{figure}

The results of this continuous setting are presented in \cref{fig:discrete_vs_continuous}, for canonical QAE as well as QAE with maximum-likelihood post-processing   \cite{Grinko2019}.
The figure clarifies the advantage of QAE variants which are not limited by the discrete resolution of canonical QAE. More explicitly, with the canonical QAE, the optimal value can be determine up to the interval $\left[0.8, 1.2\right]$, while the maximum-likelihood implementation can approach the optimum, $y^*=1$ much closer with the same effort. 

This model may be translated to the case of a discrete variable $y$.
Suppose the integer $y$ can take values up to $2^k - 1$, then it is translated to binary using $k$ qubits.
The state of these $k$ qubits, $\ket{y(\theta)}_k$ is prepared with the parameterized circuit $V(\theta)$ from \cref{fig:ry}. 
Analogously to the mapping of $x$, the value of $y$ is mapped to the expression $(x - y)^2$ via linearly controlled $R_y$ gates.
The structure of $\mathcal{A}_{\theta}$ is visualized in \cref{fig:a_operator}(b). 

The total number of qubits for $\mathcal{A}_{\theta}$ in the discrete scenario 
is $n+k+1$ versus $n+1$ in the continuous case.
Both implementations require $n$ qubits to represent $X$ and one qubit for the function
mapping and marking the states $\ket{\Psi_0}$ and $\ket{\Psi_1}$.
For the discrete case, $k$ additional qubits are used to encode the decision variable $y$.

\subsection{\label{sec:newsvendor}Newsvendor problem}

In the newsvendor problem, a newsvendor seeks the optimal amount of newspaper batches to acquire, such that the uncertain customer demand is met while as little as possible copies are left over at the end of the day \cite{Stevenson2009}.
The cost of leftover newspapers is described by overage cost and the non-realized income due to a shortage of copies by opportunity cost.
The problem of finding the optimal amount of batches to be bought is then given by
\begin{equation*}
    s^* \in \argmin_{s \in \mathbb N_0}~ \mathbb{E}[f(s, D)],
\end{equation*}
with
\begin{equation*}
    f(s, d) =
    \begin{cases}
    f_\text{opp}(s, d) = (d - s) (p_\text{sell} - p_\text{buy}) &\text{if } d \geq s \\
    f_\text{over}(s, d) = (s - d) p_\text{buy} &\text{if } d < s
    \end{cases},
\end{equation*}
where $D$ denotes the random variable representing the uncertain demand, and where each batch of newspapers is bought at a price $p_\text{buy}$ and sold
at $p_\text{sell}$.

Evaluations of the piecewise linear function $f$ can be realized with a comparison operator. 
The linear parts $f_i$, for $i \in \{\text{opp}, \text{over}\}$ are implemented with another sine approximation leveraging the linearity of $\sin^2$ around $\pi/4$, similar to the quadratic function in \cref{sec:toy}, and we refer to \cite{Woerner2018, Stamatopoulos2019} for more details.
To compute the cost function of the newsvendor, we reformulate $f$ 
as conditional addition
\begin{align*}
  f(s, d) = f_\text{over}(s, d) + \delta(d \geq s) (f_\text{opp} - f_\text{over})(s, d),
\end{align*}
where $\delta(d \geq s) = 1$ if $d \geq s$ and $0$ otherwise.
In a first step, the value $f_{\text{over}}$ is rotated onto 
the amplitude of the objective qubit.
Secondly, $\delta(d \geq s)$ is evaluated into an auxiliary
comparison qubit using the comparison operation discussed in \cref{app:qubit_comparison}.
Lastly, the second term $f_\text{opp} - f_\text{over}$ is added to the qubit amplitude as 
controlled operation on the comparison qubit. 

\begin{figure}
    \centering
    \includegraphics[width=\linewidth]{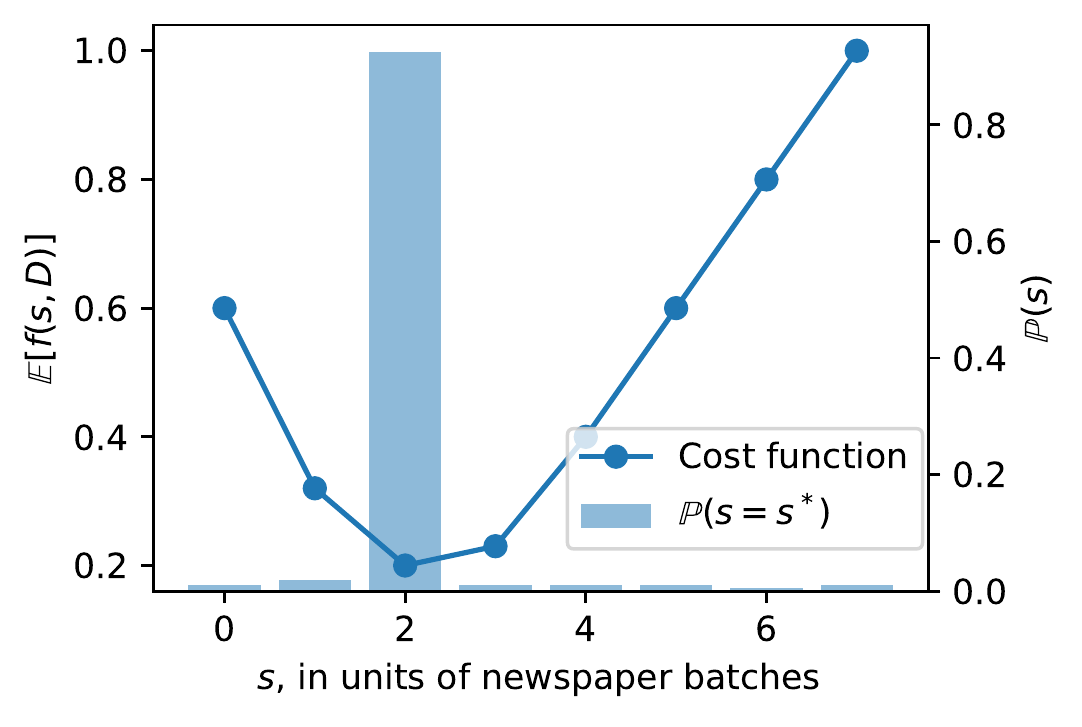}
    \caption{The newsvendor's objective function evaluated using QAE. 
    The objective function is rescaled with a factor $c = 10^{-3}$.
    }
    \label{fig:newsvendor_sampling}
\end{figure}

In this particular example, the demand is modeled with a normal distribution $D \sim \mathcal{N}(2, 1)$, represented on $n=3$ qubits and truncated to $\Omega_D = [0, 7]$.
The decision variable $s$ is encoded into $k=3$ qubits and parameterized using 
the trial state illustrated in \cref{fig:ry} with two repetitions.
Since one additional ancilla qubit for the comparison in the piece-wise linear objective is required, $\mathcal{A}$ acts in total on eight qubits.
The buy price and sell price are 0.2 and 0.5 per newspaper batch, respectively.

\Cref{fig:newsvendor_sampling} illustrates the objective function for the newsvendor problem and the result state of the optimization interpreted as 
probability density function with $\pr[s = s^*] = |\braket{s|s(\theta^*)}|^2$.
We observe that the cost function is minimal at a stock size of $s=2$ which is the stock size with the highest sampling probability of almost one.
Thus, the optimization routine successfully identified the optimal solution.

In general, the solution $\ket{s(\theta^*)}$ is in a superposition state of possible solutions $s$. During the optimization, the peak of the
probability distribution $\pr[s = s^*]$ should become increasingly 
prominent at the optimal stock value.

\subsection{\label{sec:portfolio}Portfolio optimization}

Portfolio optimization aims to select certain assets out of an available set such 
that a given objective is optimized.
Here, the considered objective function is to maximize the expected payoff while considering the buyer's willingness to take risk, modeled with the VaR.
VaR is a widely used risk metric referring to the shortfall in the $\alpha\%$ worst cases, i.e.~the $\alpha$ quantile of the expected payoff \cite{Artzner1999}. 

The return of the $k$ available assets are modelled using a multivariate random variable $X$ with sample space $\Omega_X$. 
The variable $y$ is a binary vector deciding whether an asset should be
included or not, i.e.~$y \in \{0, 1\}^k$.
Then, the expected return is given by $\mathbb{E}[y^T X]$ and the risk by 
$\mathrm{VaR}_{\alpha}(y^T X)$, as defined in \cref{app:var}. 
The resulting optimization problem then reads 
\begin{align*}
    y^* \in \argmax_{y \in \{0, 1\}^k}~ \mathbb{E}[y^T X] - q \mathrm{VaR}_{\alpha}(y^T X),
\end{align*}
where we penalize the expected return by the risk taken, weighted with a risk factor $q \in \mathbb{R}_{\geq 0}$.

Note that, while here we focus on the VaR, another 
important metric to assess the risk is the CVaR, 
or expected shortfall. 
Once the VaR is known, the CVaR can be evaluated with a consecutive run of 
QAE, see \cref{app:var}.

The expected return can be evaluated using the function mapping $F$ of 
\cref{eq:f_on_amplitudes}, if the value register contains the states $\ket{y^T x}$,
instead of $\ket{x}$.
Analogously to the linear parts in the newsvendor problem in \cref{sec:newsvendor}, we can compute $\mathbb{E}[y^T X]$ using a linear sine approximation.

\begin{figure}
    \centering
    \includegraphics[width=0.6\linewidth]{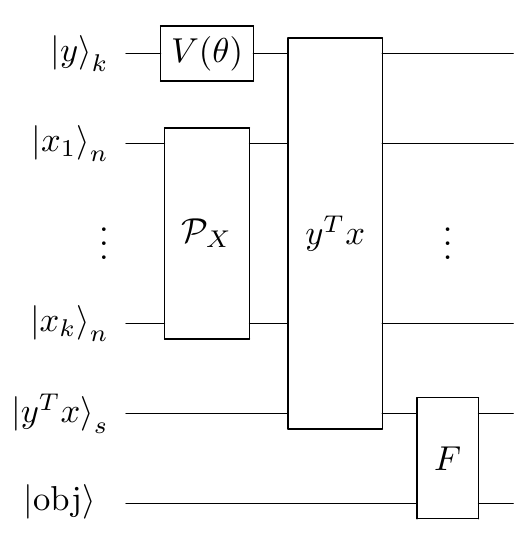}
    \caption{The circuit implementing $\mathcal{A}$ for the portfolio optimization.
             The binary decision variable $y$ is represented using $k$ qubits and the 
             multivariate distribution for the asset return is loaded into $k$ registers
             of several qubits $\ket{x_i}$. After the dot-product $y^T x$ is computed in the 
             sum-register of size $s = \log_2(k2^n)$, the objective function can be applied. For the expected return,
             $F$ implements the identity function and for the VaR, $F = F_\lambda$ is used to compute the CDF.}
    \label{fig:portfolio_circuit}
\end{figure}

To map $y^T x$ into a qubit state, i.e.~$\ket{y^T x}$, we need $s = \log_2(k 2^n)$ qubits,
which is the number of bits needed to represent the largest value the dot-product $y^T x$ can attain.
Since $y \in \{0,1\}^k$ represents a binary vector, the dot-product 
can be read as summation over all $x_i$ conditioned on $y_i$ being one
\begin{equation*}
    y^T x = \sum_{i=0}^{k-1} x_i \delta(y_i = 1).
\end{equation*}
Thus, the value $x_i$ is added to the output register as controlled
operation on qubit $\ket{y_i}$.
Known generic qubit addition operations \cite{Draper2000, Cuccaro2004}, 
\begin{equation*}
  \text{ADD} \ket a_n \ket b_n \ket 0_{n+1} = \ket a_n \ket b_n \ket{a + b}_{n+1},
\end{equation*}
can be generalized to the described controlled addition e.g. by controlling all internal operations.
The circuit to compute the expectation value is shown in \cref{fig:portfolio_circuit}.

To compute the VaR of $y^T X$, a comparison of integer values 
to the states $\ket{y^T x}$ is required -- as discussed in \cref{app:cdf,app:var}.
Thus, as for the expectation value, the dot-product circuit is employed with the function mapping $F$ not being a linear function but an integer comparator,
\begin{equation*}
    F_\lambda: \ket{y^T x}\ket{0} \mapsto 
    \begin{cases}
        \ket{y^T x}\ket{1}, \text{ if } y^T x \leq \lambda, \\
        \ket{y^T x}\ket{0}, \text{ otherwise.}
    \end{cases}
\end{equation*}
The circuit implementing the corresponding $\mathcal{A}$ is visualized in \cref{fig:portfolio_circuit}
where $F = F_\lambda$.

Suppose, we simulate portfolio optimization with $k$ assets where the
random variable for the joint distribution is modeled with $nk$ qubits.
Then, the sum $y^T x$ requires $s$ qubits to be represented and,
if \cite{Cuccaro2004} is used for the addition, $n$ ancilla qubits.
Furthermore, one qubit is needed for $F$ and, one more comparison qubit for the VaR to evaluate the CDF. 
This results in a total of $k + nk + 1$ qubits, plus $n + s (+ 1)$ ancillas, 
which is summarized  in Table~\ref{table:portfolio_qubits}.

In this application, we investigate a two-asset portfolio, $k=2$, where the joint
distribution is modeled using $nk=4$ qubits.
The returns of the assets $X$ is modeled as log-normal distribution,
$\log(X) \sim \mathcal{N}(\mu, \Sigma)$ with
\begin{equation*}
    \mu = \begin{pmatrix} 0.8 \\ 1 \protect\end{pmatrix},~
    \Sigma = \begin{pmatrix} 1 & -1 \\ -1 & 10 \protect\end{pmatrix}
\end{equation*}
and $\Omega_X = [0, 1]^2$.
The risk appetite is $q = 0.9$ and $\alpha = 0.05$.
To parameterize the binary decision variable $y$, we use 
the $R_y$ trial state, \cref{fig:ry}, with two repetitions.
The total number of qubits to compute the VaR and the expectation value is 13 and 12, respectively.

The number of qubits required for the expectation value can be reduced to 
only $k + nk + 1 = 7$ by avoiding the explicit computation of the dot-product
into an additional sum register.
Another possible implementation is to directly add the values $x_i$ into the 
linear function $F$ by additionally controlling the controlled $R_y$ rotations
on qubit $y_i$. 
This simplifies the circuit in \cref{fig:portfolio_circuit} for 
the computation of the expected returns. The computation of the VaR, however,
remains the limiting factor.

\begin{figure}
    \centering
    \includegraphics[width=\linewidth]{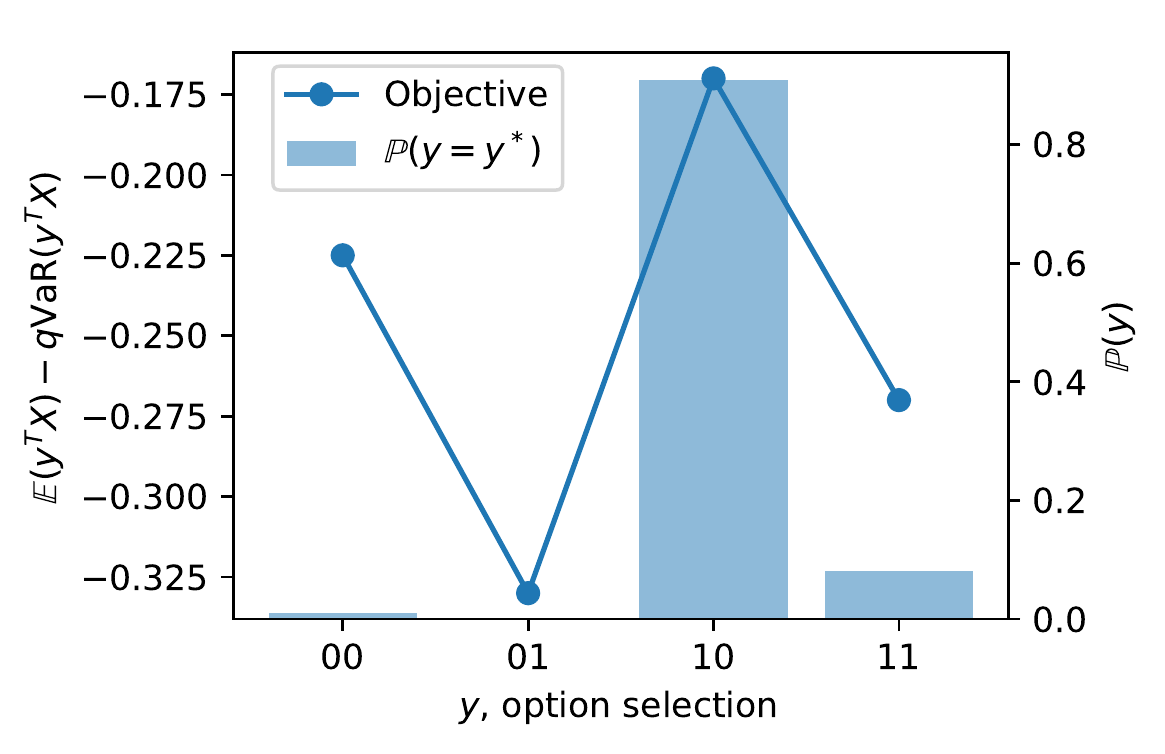}
    \caption{The portfolio optimization objective function.
             In the linear sine approximation for $\mathbb{E}[X^T y]$ a rescaling factor of $c = 0.02$ is used.}
    \label{fig:portfolio_optimized}
\end{figure}

\Cref{fig:portfolio_optimized} presents the optimization results. 
Selecting only the first asset maximizes the objective which is identified
as most likely solution by the algorithm with a probability of approximately $0.9$.

\begin{table}
    \begin{tabular}{|l|c|}
    \hline
     representation of $y$ & $k$ \\
     representation of $X$ & $nk$ \\
     marker qubit & 1 \\
     sum of asset returns & $\log_2(k 2^n)$ \\
     compare for CDF & 1 \\
     ancillas for addition & $n$ \\
     \hline
     total & $n + k + nk + \log_2(k 2^n) + 2$ \\
    \hline
    \end{tabular}
    \caption{The number of required qubits for portfolio optimization on $k$ assets, where
    each dimension of the multivariate distribution $X$ is modeled using $n$ qubits.}
    \label{table:portfolio_qubits}
\end{table}

\section{\label{sec:discussion}Conclusion and Outlook}

This paper presents a quantum algorithm for SBO, \qsbo{}, for continuous as well as discrete decision variables.
By leveraging QAE, the algorithm offers a quadratic speedup for the evaluation of the objective function compared to classical Monte Carlo simulation.
\qsbo{} is particularly suitable for objective functions that are defined as  expectation value, variance, cumulative distribution function or the (C)VaR of a random variable and functions thereof.
The application and feasibility of the algorithm is demonstrated on examples from inventory management and finance, with both, continuous and discrete decision variables.

For discrete decision variables, the algorithm evaluates the objective function for superpositions of candidate solutions depending on the chosen trial state.
For the simple examples analyzed in this paper, we always found the optimum with a high probability.
However, it requires further in-depth investigation how the chosen trial state influenced the performance and whether it is possible to choose problem-specific trial states as, e.g., in QAOA.

\section{Acknowledgements}

We acknowledge the support of  the National Centre of Competence in \textit{Research Quantum Science and Technology} (QSIT).

\appendix
\crefalias{section}{appendix}

\section{\label{app:qubit_comparison}Value comparison of qubit registers}

Let the comparison of two $n$-qubit registers $\ket{a}_n$ and $\ket{b}_n$ 
be defined as 

\begin{equation*}
    \begin{aligned}
    \ket a_n \geq \ket b_n &\Leftrightarrow
    \ket{q^{(a)}_0} \cdots \ket{q^{(a)}_{n-1}} \geq \ket{q^{(b)}_0} \cdots \ket{q^{(b)}_{n-1}} \\
    &\Leftrightarrow \sum_{i=0}^{n-1} 2^i q^{(a)}_i \geq \sum_{i=0}^{n-1} 2^i q^{(b)}_i.
    \end{aligned}
\end{equation*}

The idea for the implementation of the comparison is to reformulate the
statement $a \geq b$ for two $n$-bit integers $a$ and $b$ to $a + (2^n - b) \geq 2^n$.
The sum on the left hand side, represented with $n+1$ bits, is
larger or equal to $2^n$ exactly if the most-significant bit is 1.
Thus, the comparison can be broken down into two steps.
First we calculate the $(2^n - b)$ into a qubit register and then add $a$. 
Finally the most-significant bit of $a + (2^n - b)$ is returned.
Note that these additions can be done in-place, where the input is overwritten with the 
output to use less qubits.

\begin{figure}
    \centering
    \includegraphics[width=0.8\linewidth]{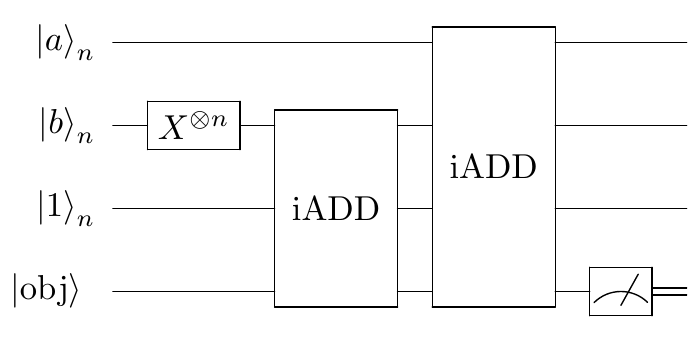}
    \caption{The circuit computing $a \geq b$. 
             The first two blocks, $X^{\otimes n}$ and iADD, compute the two's complement
             of $b$, where iADD denots the in-place addition, i.e., the value of the second register is overwritten with the sum. 
             The second addition adds the value of $a$. The carry qubit is then
             contains the result and is set to 1 if $a \geq b$, otherwise 0.
             }
    \label{fig:comparator_circuit}
\end{figure}

Computing $\ket{2^n - b}_{n+1}$ from $\ket{b}_n$ can be done using classical binary logic.
We first flip all qubits of $b$ by applying NOT gates and then use an adder circuit
to add the value 1.
By using the adder introduced in \cite{Cuccaro2004} in-place addition can be used 
so the state $(2^n - b)$ is stored in the register of $\ket{b}_n$ plus a carry qubit 
for the most-significant bit. 
Then, add $\ket{a}_n$ into the register of $\ket{b}_n$ and the carry qubit to
obtain $(2^n - b) + a$. As the addition circuit require both sum terms to have
the same number of bits, $\ket{a}_n$ must be padded with an extra ancilla qubit
in state $\ket{0}$ acting as most-significant bit.
The carry qubit is now in state $\ket{1}$ is $a \geq b$, otherwise in state $\ket{0}$
and can be measured.
These operations are schematically visualized in \cref{fig:comparator_circuit}.

Comparing the value a single register $\ket{a}_n$ to a fixed integer value $b$ 
instead of an integer encoded into a qubit register $\ket{b}_n$ follows the same procedure. 
However, $(2^n - b)$ can be computed classically and the controlled operations 
to add $a$ and $(2^n - b)$ be replaced by classical controls.

\section{\label{app:cdf}Cumulative Distribution Function}

The CDF of a random variable $X$ for a value $x$ is
$\pr[X \leq x]$, i.e.~the probability to sample a value smaller than, or equal to, $x$.
For a discretized $X$, taking values $\Omega_X = \{x_i\}_{i=0}^{N-1}$, $N = 2^n$ with 
corresponding probabilities $p_{x_i}$, the CDF can be computed by summing the 
probabilities of all samples $x_i \leq x$, i.e.
\begin{equation}
    \pr[X \leq x] = \sum_{x_i \leq x} p_{x_i} = \sum_{\hat{x}=0}^{\lambda} p_{\phi(\hat{x})},
    \label{eq:cdf}
\end{equation}
where $\phi: \{0, \ldots, N - 1\} \rightarrow \Omega_X$ and 
$\lambda = \phi^{-1}(x)$.

The $\mathcal{A}$ operator encoding the sum \cref{eq:cdf} is 
constructed analogously to the expectation value of \cref{sec:preliminaries},
i.e. $\mathcal{A} = F(\mathcal{P}_X \otimes \id)$. 
Instead of a linear or quadratic function, however, $F$ implements
a step-function, that flips the state of the last qubit if $\hat{x} \leq \lambda$.
This $F = F_\lambda$ equals the comparison operator introduced in \cref{app:qubit_comparison},
where the value we compare to is the integer $\lambda$.
Explicitly, $\mathcal{A}$ reads
\begin{equation*}
    \begin{aligned}
    \mathcal{A} \ket{0}_n \ket{0} &= \sum_{\hat{x}=0}^{\lambda} \sqrt{p_{\phi(\hat{x})}} \ket{\hat{x}}_n \ket{1} \\
                                 &+ \sum_{\hat{x}=\lambda+1}^{N - 1} \sqrt{p_{\phi(\hat{x})}} \ket{\hat{x}}_n \ket{0},
    \end{aligned}
\end{equation*}
as described in \cite{Woerner2018}.
The state $\ket{\Psi_1}$, of which the amplitude is estimated, is defined as all states where 
the last qubit is $\ket{1}$.

\section{\label{app:var}(Conditional) Value at Risk}
The VaR at level $\alpha \in [0,1]$ of a random variable $X$ is the smallest value
$x_\alpha$ such that the CDF is larger than $\alpha$ \cite{Artzner1999}.
Let $\Omega_X \subset \mathbb R$ be the sample space of $X$, then
\begin{equation*}
  \mathrm{VaR}_\alpha(X) = x_\alpha := \min\left\{ x \in \Omega_X : \pr[X \leq x)]
                                                          \geq \alpha \right\}.
\end{equation*}
Since we know how to compute the CDF of $X$, for a value $x$
with QAE the
VaR can easily be computed by finding the root of the function 
$g(x) = \pr[X \leq x] - \alpha \stackrel{!}{=} 0$.
This function is monotone and intersects with 0, therefore a bisection search 
is a suitable method to find the root.

The CVaR at level $\alpha \in [0,1]$ is the expectation
value of $X$ restricted to values smaller equals $x_\alpha$, i.e.
\begin{equation*}
    \mathrm{CVaR}_{\alpha}(X) = \mathbb{E}[X \vert X \leq x_\alpha]. 
\end{equation*}
To estimate this quantity with QAE, we can use the same techniques introduced 
in \cref{sec:preliminaries} and \cref{sec:newsvendor} to compute the expectation value
of $\mathbb{E}[X]$. However, we additionally must
restrict the sampled values of $X$ to be smaller or equal to the VaR.
This can be achieved by first applying a qubit comparison to flag all states $\leq \lambda_\alpha$, $\lambda_\alpha = \phi^{-1}(x_\alpha)$, and then compute the expectation value on this sub-state, 
where $\phi$ is defined in \cref{app:cdf}.

The action of $\mathcal{A}$ can thus be written as 
\begin{equation*}
    \begin{aligned}
    &\mathcal A\ket{0}_{n}\ket{0} =
    \left(\sum_{\hat{x}=0}^{\lambda_\alpha} \sqrt{p_{\phi(\hat{x})} \frac{\hat{x}}{\lambda_\alpha}} \ket{\hat{x}}_n\right)\ket 1 \\
    &+ \left(\sum_{\hat{x}=0}^{\lambda_\alpha} \sqrt{p_{\phi(\hat{x})} \left(1 - \frac{\hat{x}}{\lambda_\alpha}\right)} \ket{\hat{x}}_n +
             \sum_{\hat{x}=\lambda_\alpha+1}^{2^n - 1} \sqrt{p_{\phi(\hat{x})}} \ket{\hat{x}}_n\right)\ket 0
    \end{aligned}
\end{equation*}
where the factor $\lambda_\alpha^{-1}$ is used to scale $\hat{x}$ to $[0, 1]$.
As the expectation value is not taken over the full domain $\Omega_X$, but only over
the subset where $X \leq x_\alpha$, we finally need to renormalize the 
probabilities to 1 via the factor $\pr[X \leq x_\alpha]^{-1}$,
\begin{align*}
    \mathrm{CVaR}_\alpha(X) = \frac{\tilde{a} \lambda_\alpha}{\pr[X \leq x_\alpha]} = \frac{\lambda_\alpha}{\pr[X \leq x_\alpha]} \sum_{x=0}^{\lambda_\alpha} \frac{x}{\lambda_\alpha} p_{\phi(x)}.
\end{align*}

\bibliographystyle{IEEEtranN}
\bibliography{references}

\end{document}